\documentclass{rmaa}

\input epsf.tex
\usepackage{longtable}

\newcommand{\degr}{\mbox{$^{\rm o}$}}

\title{Optical Coordinates of Southern Planetary 
Nebulae}
\author{S. Kimeswenger\altaffilmark{1}
  \affil{Institut f{\"u}r Astrophysik, Universit{\"a}t Innsbruck, Austria.} 
}

\altaffiltext{1}{Email: Stefan.Kimeswenger@uibk.ac.at}

\fulladdresses{
\item S. Kimeswenger: Institut f\"ur Astrophysik der Leopold--Franzens Universit\"at Innsbruck, Technikerst. 25, A-6020 Innsbruck, Austria (Stefan.Kimeswenger@uibk.ac.at).
}

\shortauthor{KIMESWENGER}
\shorttitle{Optical Coordinates of PNe - southern objects}

\SetVolume{37} \SetFirstPage{-999} \SetYear{2001}
\ReceivedDate{2000 December 29} 
\AcceptedDate{2001 May 3} 

\resumen{
Un conjunto homog{\'e}neo de nuevas medidas de casi todos los (995 sobre 1007) objetos del cat{\'a}logo de
Strasbourg-ESO de Nebulae planetario gal{\'a}ctico o en el primer suplemento de este cat{\'a}logo en el {\'a}rea
cubri{\'o} por el DENIS survey ($\delta < +2\degr$) se da aqu{\'{\i}}. 24  nuevos y 27 falsas identificaciones
cruzadas con fuentes en el IRAS PSC catalogan y una cierta confusi{\'o}n en la literatura se enumera
tambi{\'e}n.}
\abstract{A homogeneous set of 
new measurements of nearly all (995 out of 1007)
objects from the 
{\it Strasbourg-ESO
Catalogue of Galactic Planetary Nebulae}
or in the first supplement of this catalogue
in the area covered by the 
DENIS survey ($\delta < +2\degr$) is given here. 24 new and 27 wrong 
crossidentifications with sources in the IRAS PSC catalogue and some 
confusion in the literature is listed as well.}
      

\listofauthors{S. Kimeswenger}
\indexauthor{Kimeswenger, S.}
\begin{document}

\maketitle

\section{Introduction}
\label{sec:Introduction}

In the preparation course of a of catalogues of near infrared (NIR) sources of the 
DENIS survey (Epchtein et al. 1994, 1997) and
during the measurements of planetary nebulae (Kimeswenger et al. 1998a), 
we found that the optical coordinates of a significant 
portion of the known galactic planetary nebulae (PNe) are inaccurate. 
These inaccuracies impair the use of automated cross-identifications for, e.g.,
colour-colour diagrams (Kimeswenger 1997). Also 
a wealth of problems arise in electronic data bases (Kimeswenger et al. 1998b). 
The coordinates of the objects in the large PNe--catalogues 
(e.g. Perek \& Kohoutek 1967 = CGPN,  
Acker et al. 1992a = SECGPN, Acker et al. 1996) 
mostly were taken from the original literature. 
Even the coordinates in the IAC Morphological Catalog of Northern Galactic PNe
 (Manchado et al. 1996) are not derived from the images, but copied from the
SECGPN only. Thus this catalogue can not be seen as northern counterpart to this
compilation here.
More accurate optical coordinates of a small set of nebulae were re--determined later (Milne 
1973, 1976; Blackwell \& Purton 1981). 
The accuracy of the recently detected objects  is in most cases rather good.
However 
older coordinates are often given $\pm$0{$.\!$\arcmin}1 or 
worse. Also, a considerable number of central stars (CSPN) of extended nebulae 
were identified in the 
meanwhile. Those nebula got a better defined ''center''.
A low accuracy is insufficient for satellite observations or 
those of the 
new generation very large telescopes. Even 
the identification on the crowded  12\arcmin$\times$12\arcmin\ fields coming from the 
DENIS survey 
was often quite difficult  (Epchtein et al. 1994). We measured all coordinates by means of the Digital Sky 
Survey and some supplementary material, like CCD digitisation from the Schmidt 
plate copies for faint nebulae or I band images obtained by the DENIS survey.
Although there exist for a small number of 
nebulae more accurate radio coordinates, we measured those nebulae too
and thus got a highly homogeneous data set of optical coordinates.

\section{Nomenclature}
The name (identifier)  of the nebula is usually formed using the galactic 
coordinates (truncated at 1/10$^{th}$ of a degree)
\[ \mbox{PN G}LLL.l{\pm}BB.b \]
\noindent as it was proposed for PNe by Acker et al. (1992a).
A change of coordinates would imply a change of the name identifiers.
According to the recommendations of the IAU 
(Lortet \& Spite 1986, Dickel et al. 1987),
and in order to avoid severe confusion,
no changes of the identifier should be
applied later.
This procedure is also used in data bases like, e.g., SIMBAD.
The used identifiers are just starting to get ambiguous. The current 
lists (true PNe only) have not a real problem up to now 
(one case were a letter had to be added). This was also done at other data bases too:
e.g. the IRAS data base. But the situation differs. There are no entries
added to the IRAS data base anymore. But the currently running surveys for 
emission line objects in general and PNe (e.g. Parker et al. 1999, Beaulieu et al. 1999, 
van de Steene \& Jacoby 1999, Beaulieu 2000) 
will add a few hundred sources in the bulge
region within the next few years. Thus, and as the accuracy of coordinates is now significantly
better than 0.01 degrees
($\equiv$36\arcsec), we suggest the use of 
\[ \mbox{GPN G}LLL.ll{\pm}BB.bb \]
\noindent in the future.
The descriptor GPN stands for galactic planetary nebula. 
 The
remaining string is formed as described in the dictionary of nomenclature
cited above.
This also fulfills the recommendations of the IAU
Dictonary of Nomenclature, where a list identifier should contain at least 3 characters 
(the 'G' after the blanc does not count as it is part of the coordinate) and that the coordinate should be 
choosen not to the limit to be unambiguous, but to that given by the expected coordinate accuracy (worst case).

\section{The Data}
A 5\arcmin$\times$5\arcmin\ field at the nominal position as given in the SECGPN of each 
source was loaded from the 
Digital Sky Survey (DSS) . These maps were compared, as far as possible, 
with the finding charts of the 
original discoveries and with those given in the SECGPN and the CGPN.
The coordinates of these plates were 
calibrated by means of the Guide Star Catalogue (GSC). This yields  
coordinates better than the pixel resolution (relative to the GSC reference frame). 
In case of compact nebulae and extended nebulae, where the CSPN could be identified on the 
DSS scans (the vast majority of the objects),  a gaussian fit was used to derive 
the position. If an object was too faint on the DSS scans, a CCD scan of the 
Schmidt plate copy (by means of a microscope system) were obtained. In a few cases the images of 
Schwarz et al. (1992) were taken. In one case (A 58) the image in Pollacco et al.
(1992) an in Kimeswenger et al. (2000) and in another case (NGC 5189) that one in Kimeswenger et al. (1998a) 
were used. In case of about 15 recently discovered nebulae (Jacoby \& van de
Steene - listed in Acker et al. 1996), NIR finding charts were
available only, I band images of the DENIS survey were used.
The positions of these auxiliary frames were calibrated by 
use of the DSS scans. This method gives an overall accuracy $<$ 1\arcsec, relative to the GSC stars. 
In case of the small number of very extended 
irregular nebulae without a known CSPN the geometric centre (of the smallest circle 
embedding the object) was chosen. The coordinates of all objects are given 
in Tab. 1.
Also remarks corresponding to misidentifications on the finding 
charts (FC) in Acker et al. (1992a) or Acker et al. (1996) are given as 
supplementary information. For 12 nebulae  no
identification was possible (listed in Tab. 2).
\begin{figure}
\centerline{\epsfxsize=88mm \epsfbox{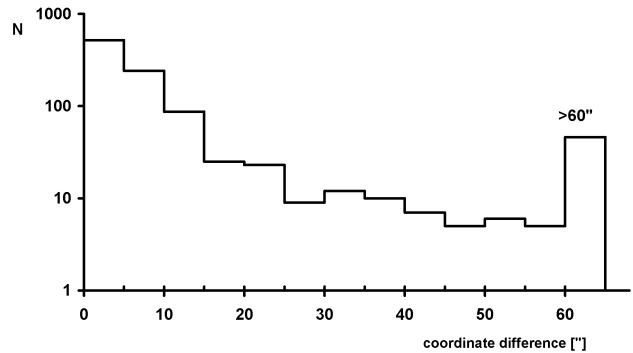}}
\caption{Frequency histogram of the positional difference beween the
 determination in 
this work and the positions given in the SECGPN and in Acker et al. (1996).}
\end{figure}
A set of objects (88) has multiple records and misclassifications from the literature,
mostly ''galaxies'' from sky survey plate searches,
in the SIMBAD data base (current status). 
Those are listed in Tab.1 too. In some not so obvious cases they are 
described in detail in section 5.
Fig. 1 gives a frequency histogram of coordinate offsets between
those listed in the catalogues and those measured here. 
As can be seen from the diagram, the new determination of coordinates 
led to a marked improvement for a significant fraction of PNe.

\section{Crossidentifications}
\subsection{IRAS}
The existing crossidentification with IRAS sources were checked by means of the 
coordinates and the error ellipse of the IRAS PSC (which corresponds to a 2 $\sigma$ 
correlation). In case of extended sources the outermost region of a nebula was used 
for the boundary condition. The coordinate information and the size of the IRAS error
ellipse were used mainly for the decision. Each rejection was then investigated 
individually on basis of IR colours, size 

\onecolumn
\begin{longtable}{l l l l r r r r l}
\caption{Coordinates and crossidentifications for southern PNe}\\
\hline\hline
\tt\tiny
GPN & \tt\tiny PN G & \tt\tiny usual name & \tt\tiny IRAS & \tt\tiny $\alpha$(J2000.0) & \tt\tiny $\delta$(J2000.0) & \tt\tiny $\ell$\phantom{XXX} & \tt\tiny
$b$\phantom{XXX} & \tt\tiny
remark(s) and other \vspace{-2.5mm} \\
& \tt\tiny (this work) & \tt\tiny (Acker et al.1992a, 1996) & \tt\tiny \phantom{17255$-$2448B} & & & \tt\tiny \phantom{359.999} & \tt\tiny \phantom{-99.999} & \tt\tiny sources associated \phantom{XXXXXXXXXXXXX}\\
\hline
\endfirsthead
\caption{continued}\\
\hline\hline
\tt\tiny
GPN & \tt\tiny PN G & \tt\tiny usual name & \tt\tiny IRAS & \tt\tiny $\alpha$(J2000.0) & \tt\tiny $\delta$(J2000.0) & \tt\tiny $\ell$\phantom{XXX} & \tt\tiny
$b$\phantom{XXX} & \tt\tiny
remark(s) and other \vspace{-2.5mm} \\
& \tt\tiny (this work) & \tt\tiny (Acker et al.1992a, 1996) & \tt\tiny \phantom{17255$-$2448B} & & & \tt\tiny \phantom{359.999} & \tt\tiny \phantom{-99.999} & \tt\tiny sources associated \phantom{XXXXXXXXXXXXX} \\
\hline \vspace{-2.5mm}
\endhead
\hline\hline \vspace{-2.5mm}
\endfoot
\tt\tiny 000.00$-$06.85 & \tt\tiny  000.0$-$06.8 & \tt\tiny  H 1$-$62             & \tt\tiny   18100$-$3220 & \tt\tiny  18 13 18.1 & \tt\tiny  $-$32 19 43 & \tt\tiny    0.000 & \tt\tiny   $-$6.851 & \tt\tiny   \vspace{-2.5mm} \\
\tt\tiny 000.03$+$02.06 & \tt\tiny  000.0$+$02.0 & \tt\tiny  K 6$-$4              & \tt\tiny   17345$-$2747 & \tt\tiny  17 37 43.0 & \tt\tiny  $-$27 49 06 & \tt\tiny    0.031 & \tt\tiny    2.066 & \tt\tiny   \vspace{-2.5mm} \\
\tt\tiny 000.10$-$01.91 & \tt\tiny  000.1$-$01.9 & \tt\tiny  JaSt 90            & \tt\tiny              & \tt\tiny  17 53 24.4 & \tt\tiny  $-$29 49 48 & \tt\tiny    0.105 & \tt\tiny   $-$1.915 & \tt\tiny   \vspace{-2.5mm} \\
\tt\tiny 000.12$-$01.14 & \tt\tiny  000.1$-$01.1 & \tt\tiny  M 3$-$43             & \tt\tiny   17472$-$2924 & \tt\tiny  17 50 24.2 & \tt\tiny  $-$29 25 20 & \tt\tiny    0.123 & \tt\tiny   $-$1.145 & \tt\tiny   \vspace{-2.5mm} \\
\tt\tiny 000.12$-$08.09 & \tt\tiny  000.1$-$08.0 & \tt\tiny  SB 1               & \tt\tiny              & \tt\tiny  18 18 48.9 & \tt\tiny  $-$32 47 57 & \tt\tiny    0.120 & \tt\tiny   $-$8.099 & \tt\tiny   \vspace{-2.5mm} \\
\tt\tiny 000.13$+$02.68 & \tt\tiny  000.1$+$02.6 & \tt\tiny  Al 2$-$J             & \tt\tiny              & \tt\tiny  17 35 35.1 & \tt\tiny  $-$27 24 02 & \tt\tiny    0.131 & \tt\tiny    2.689 & \tt\tiny   \vspace{-2.5mm} \\
\tt\tiny 000.15$-$05.61 & \tt\tiny  000.1$-$05.6 & \tt\tiny  H 2$-$40             & \tt\tiny              & \tt\tiny  18 08 30.8 & \tt\tiny  $-$31 36 31 & \tt\tiny    0.156 & \tt\tiny   $-$5.616 & \tt\tiny   \vspace{-2.5mm} \\
\tt\tiny 000.16$+$17.24 & \tt\tiny  000.1$+$17.2 & \tt\tiny  PC 12              & \tt\tiny   16409$-$1851 & \tt\tiny  16 43 53.8 & \tt\tiny  $-$18 57 13 & \tt\tiny    0.166 & \tt\tiny   17.248 & \tt\tiny  IRAS 1640$-$188P04 \vspace{-2.5mm} \\
\end{longtable}
{\tt\tiny \vspace{-3mm}
\begin{tabular}{l l}
$[1]$ &  FC in SECGPN rotated by 180\degr \\
$[2]$ & FC(SECGPN) wrong sky region\\
$[3]$ & wrong CSPN in FC(SECGPN) - real one 7\arcsec\ E of indicated bright star\\
$[4]$ & wrong CSPN in FC(SECGPN) - real one 18\arcsec\ NEE of indicated star\\
$[5]$ & FC(SECGPN) reversed in $x$\\
$[6]$ & wrong CSPN in FC(SECGPN) - real one faint star SE fo marked one\\
$[7]$ &  FC(SECGPN) wrong - real PN 2.4\arcmin\ W, 1\arcmin\ N of indicated star\\
$[8]$ & FC(SECGPN) wrong identifier at chart\\
$[9]$ & FC(SECGPN) rotated 90\degr\ conter-clock\\
$[10]$ & FC(SECGPN) maked only SE filament of very big nebula\\
$[11]$ & FC(SECGPN) reversed in $y$\\
$[12]$ & detailed remarks in section 5 \\
$[13]$ & FC(Acker et al. 1996) - wrong object indicated on FC - real object 20\arcsec\ SE\\
$[14]$ & FC(Acker et al. 1996) - wrong CSPN indicated on FC - real CSPN 10\arcsec\ E\\
$[15]$ & FC(Acker et al. 1996) - wrong scale beside chart\\
$[16]$ & FC(Acker et al. 1996) reversed in $x$\\
$[17]$ & FC(Acker et al. 1996) - dss chart wrong sky region, small stamp reversed in $x$\\
$[18]$ & FC(Acker et al. 1996) - dss chart wrong sky region, small stamp reversed in $y$\\
\end{tabular}
}

\twocolumn

\noindent and morphology of the nebula.
In a few cases  the  
coordinates are within the errors, but an other 
bright object 
(usually a star) is more likely the correct identification. 27 (4.5 \%) objects 
listed as IRAS PSC counterparts in Acker et al. (1992, 1996) were 
rejected to be the proper optical counterpart to the IRAS source.
Sources without an identification with an IR source were checked 
with the IRAS--PSC on basis of coordinates and IR colours. This results in 24
new identifications. The identifications assumed to
be correct are listed in Tab. 1 and Tab 2.
The rejected objects are 
given in Tab. 3. 
In Fig. 2 the frequency histogram of the differences of the source position between the 
IRAS PSC source and the coordinates measured here and that of the object in
the SECGPN to the IRAS source is shown.  

\begin{figure}
\centerline{\epsfxsize=88mm \epsfbox{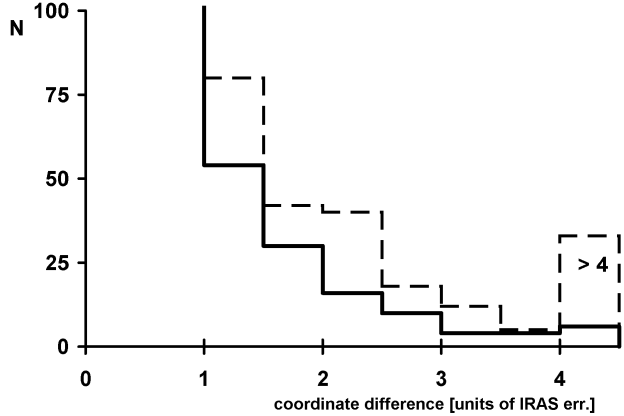}}
\caption{Frequency histogram of the positional difference beween the 
determination in 
this work and the positions of the identified IRAS PSC sources (solid line)
and those of 
SECGPN positions to the same IRAS PSC sources after removing the rejected
sources (dashed line)}
\end{figure}

\begin{table}
\caption{Objects not identified on plates or images}
\phantom{XX}

\centerline{\tiny
\begin{tabular}{ l l l l }
\hline\hline
GPN & PN G & usual name & 
remark(s) \\
\hline
000.17$-$01.21 & 000.1$-$01.2 & JaSt 72 & IRAS 17476$-$2923\\
000.33$-$01.64 & 000.3$-$01.6 & JaSt 83 & \\
001.02$+$01.35 & 001.0$+$01.3 & JaSt 39 & \\
001.58$+$01.51 & 001.5$+$01.5 & JaSt 43 & \\
021.66$+$00.81 & 021.6$+$00.8 & VSP 2$-$19 & IRAS 18250$-$0940  \\
292.97$+$01.99 & 292.7$+$01.9 & Wray 16-93 & IRAS 11285$-$5900  \\
354.43$+$04.03 & 354.4$+$04.0 & Te 233 & \\
358.44$+$01.66 & 358.4$+$01.6 & JaSt 3 & \\
358.63$+$00.75 & 358.6$+$00.7 & JaSt 16 & most likely asociated with\\
& & & GPSR 358.639$+$0.757 \\
358.83$+$01.78 & 358.8$+$01.7 & JaSt 5 &\\
359.02$+$01.16 & 359.0$+$01.1 & JaSt 9 & \\
359.23$+$01.36 & 359.2$+$01.3 & JaSt 8 & \\
\hline\hline
\end{tabular}}
\end{table}

\begin{table}
\caption{IRAS sources rejected to be associated with PNe from SECGPN or Acker et al. (1996).}
\phantom{XX}

\centerline{\tt\tiny
\begin{tabular}{l l}
\hline\hline
IRAS & remark(s) \\
\hline
  17322-2721 & clearly too far away \\
  17519-2957 & clearly too far away \\
  17548-2944 & clearly too far away \\
  18152-3156 & clearly too far away \\
  17374-2700 & clearly too far away \\
  17371-2641 & clearly too far away \\
  18131-3008 & bright star beside it - IRAS source also has stellar colours \phantom{XXXX} \\
  18001-2659  & clearly too far away and IRAS source has stellar colours \\ 
  18092-2752  & about 2 error ellipses away, stellar colours and \\
               &          bright stars at IRAS PSC position \\
  18027-2630  & clearly too far away \\  
  18150-2550  & clearly too far away \\
  18114-2443  & clearly too far away \\  
  18335-2151  & clearly too far away \\
  18265-1908  & clearly too far away \\ 
  07415-3435  & clearly too far away and bright star at exact position \\
  12202-6344   & clearly too far away \\
  15478-4817   & clearly too far away \\
  17023-5226   & clearly too far away \\
  17012-3049   & clearly too far away and clearly stellar colours \\
  17484-3135 & Al2-O mixed with Al2-P -- see section 5 \\
  17253-2824  & clearly too far away \\
  17236-2739 & mixed with SaWe2 and PBOZ 4 -- see section 5\\
  17471-3034  & clearly too far away \\
  17329-2756  & clearly too far away \\
  18115-3237  & clearly too far away \\
  17191-4700  & 2.5 error ellipses, stellar [12]-[25] and \\
                         & exact position of the bright star SE of the nebula.\\                      
  18038-3603  & 3.1 error ellipses, stellar colours and exact coordinates \\
                         & of star beside \\
\hline\hline
\end{tabular}}
\end{table}

\section{Individual Objects}

Some individual objects erroneously classified or identified as PNe or being doubtful PNe are separately
discussed as follows: 
\begin{description}
\item[]{\tt GPN G254.67$+$00.21 = PN G254.6$+$00.2 = Ns 238} :\\Neither this object has the morphology of a PN,
nor are the IRAS colours and the total FIR flux suitable for a PN. Thus it was included 
in a sample of star forming regions by Chan et al. (1996). Also the NIR photometry
with a wide beam (Liseau et al. 1992) is more likely that of a star forming region (Kimeswenger \& Weinberger 2001). 
Rosseau \& Perie (1996)
list four stars in that region (VdBH 13aA, 13 aB, 13b and 13c) as ''embedded in a reflection nebula''. 
The survey of
OH emission towards post--AGB stars (teLintel Hekkert \& Chapman 1996)
led to no detection, while the region appears rather bright in CO (Wouterloot \& Brand 1989).
Goebel et al (1995) select this source as an IRAS LRS possible carbon star having some
unusual features around 11$\mu$m. The IRAS source is not well associated with the
nebula and the brightest star, which is classified as a K giant, is somewhat displaced towards west.
Inspecting the images of Schwarz et al. (1992), I found that there is a lot of H$\alpha$ emission
towards the location of the IRAS source and an indication of a bipolar structure, while {\tt [OIII]} 
emission is only around the stars cited above towards the edge of the system.
Thus the nature of this object has to be questioned.
\item[]{\tt GPN G261.15$+$00.24 = PN G260.7$+$00.9 = Vo3 = PN G260.1$+$00.2} :\\
This object was identified by Volk  in a 
spectroscopic survey of IRAS sources (priv. comm. to Acker) as a planetary nebula. 
The finding chart in the SECGPN is correct. Also the crossidentification to
IRAS 08355-4027 is correct. The nebula is misidentified in Acker et al. (1992a) as one 
given in Holmberg et al. (1978): there it is identical to BRAN 174
(Brand et al. 1986) and identified as an HII region around the emission line stars 
ESO-HA 161 and
ESO-HA 162 (Pettersson \& Reipurth 1994).
The real PN is identical to BRAN 163 and was ''rediscovered'' as PN 0835-4027
by
van de Steene \& Pottasch (1993) and thus the identifier PN G260.1+00.2 was 
assigned (Acker et al., 1996).
\item[]{\tt GPN G272.40$-$05.96 = PN G272.4$-$05.9 = MeWe 1$-$1 = Bran 199 = ESO 165-6} :\\ 
Feitzinger \& Galinski (1985)
identify this object as
elliptical galaxy with many stars superimposed. The ESO Uppsala survey remarks: ''possible
neb.''.
The morphology and spectroscopy (SECGPN) indicates clearly the nature as a PN.
\item[]{{\tt GPN G358.01$-$02.73 = PN G358.3$+$02.6} and {\tt GPN G358.39$-$02.50 = PN G358.5$-$02.5}} :\\ 
Those nebula are reversed in the SECGPN.
While the finding charts are correctly given as in Allen (1979), the identification 
was wrong (just the opposite). This leads to some confusion with the previously 
defined PK identifiers in the data bases.
\item[]{\tt GPN G358.88$+$00.05 = PN G358.8$-$00.0} :\\ This nebula is attached to the identifier
 Terz N 2022 in the SECGPN and in SIMBAD (Terzan \& Ounnas 1988)
 In the data base at the same coordinates 
a nebula named Terz N 124 is given - the origin for this identifier might be the
work of Acker et al. (1992b) were it is given as TeGo 124.
There are no finding charts available. But as the coordinates are 
identical to arcseconds, it is assumed that this is the same source. The coordinates of the 
IRAS source 17395$-$2950 are identical too. This source is extremely cold according to the FIR. 
\item[]{\tt GPN G011.39$+$17.98 = PN G011.4$+$17.9 = PK 011+18 1 = PK 011+17 1 } :\\ 
This object is labeled in the supplement to the SECGPN
as DHW 2. There exist two discovery lists of Dengel, Hartl and Weinberger
(Dengel et al. 1979, 1980)
The more commonly known second one has the identifiers DHW
(and sometimes DeHt). Thus
we use
the identifier DHW 1-2 (similar to recent the recent work by  Saurer et al. 1997) for this 
object to avoid overlap to the
second object in the second list. In SIMBAD two independent entries
at the same coordinates are given.
\item[]{\tt KeWe 1 = KW 11} :\\ The nebula KeWe 1 (Kerber \& Claeskens 1997) is identical to that
one indicated as KW 11 (Kerber \& Weinberger 1995).
KW 1, as used in Acker et al (1996), has nothing to do
with KW 1 = IRAS 00592+6530 = KLW 9 used in SIMBAD.
\item[]{\tt KeWe n = KW n n=2,3,4,5} :\\
The nebulae KeWe 2, 3, 4 and 5 (Kerber et al. 1998) are identical to
that one indicated as KW 2, 3, 4 and 5 in SIMBAD and identical to KW 6, 8, 9 and 13 in
Kerber \& Weinberger (1995).
\end{description}

\section{Conclusion}

The considerable improvement of coordinates and elimination of
misidentifications presented in this paper are important steps in refining 
the existing catalogues of galactic planetary nebulae. This gives a 
powerful tool for automated and semiautomatic crossidentification for forthcoming 
surveys in other wavelengths and for the observations with space--born equipment or 
large telescopes of the next generation. 
%The resulting corrections of the existing 
%crossidentifications and the improvement of coordinate differences (see Fig. 2) show, 
%the improvement by using the Digitised Sky Survey for positional measurements.

\acknowledgments

This  project  was  supported by the FWF  projects  P8700-PHY,
P10036-AST and P11675-AST.
The Digitised Sky Survey is based on photographic data obtained using the UK Schmidt Telescope. The UK Schmidt
      Telescope was operated by the Royal Observatory Edinburgh, with funding from the UK
      Science and Engineering Research Council, until 1988 June, and thereafter by the
      Anglo-Australian Observatory. Original plate material is copyright © the Royal Observatory
      Edinburgh and the Anglo-Australian Observatory. The plates were processed into the present
      compressed digital form with their permission. The Digitized Sky Survey was produced at the
      Space Telescope Science Institute under US Government grant NAG W-2166.

\end{document}